\documentclass[preprint]{aastex}
\slugcomment{accepted by Icarus, June 23, 2004}

\def \WW             {{1998~WW$_{31}$}}

\def \QT             {{2001~QT$_{297}$}}

\def \RZ             {{(66652) 1999~RZ$_{253}$}}
\def \ref            {\noindent\hangindent0.5in\hangafter=1}

\begin{document}

\title  {The Orbit, Mass, and Albedo of Transneptunian Binary \RZ .}

\author{ Keith S.~Noll, Denise C.~Stephens} 
\affil{$^1$ Space Telescope Science Institute, Baltimore, MD 21218}
\email{{\bf Email: } noll@stsci.edu}

\author{ Will M.~Grundy}
\affil{Lowell Observatory, 1400 W.~Mars Hill Rd., Flagstaff, AZ 86001}

\author{ Ian Griffin}
\affil{Museum of Science and Industry, Manchester, UK}

{\bf Proposed Running Head:} The Orbit of TNO Binary (66652) 1999 RZ$_{253}$

\bigskip
{\bf Editorial Correspondence:} Dr. Keith S. Noll

Space Telescope Science Institute

3700 San Martin Drive

Baltimore, MD 21218

phone: 410-338-1828

fax: 410-338-4767

email: noll@stsci.edu

\bigskip
Pages: 15

Tables: 4

Figures: 4

\newpage

\centerline{\bf ABSTRACT  }

\noindent{We have observed \RZ\ with the Hubble Space Telescope at
seven separate epochs and have fit an orbit to the observed relative
positions of this binary.  Two orbital solutions have been identified
that differ primarily in the inclination of the orbit plane.  The best
fit corresponds to an orbital period, P=46.263$\pm {0.006\atop 0.074}$
days, semimajor axis a=4,660$\pm$170 km and  orbital eccentricity
e=0.460$\pm$0.013 corresponding to a system mass m=3.7$\pm0.4 \times
10^{18}$ kg.  For a density of $\rho$ = 1000 kg m$^{-3}$ the albedo at
477 nm is $p_{477}$= 0.12$\pm 0.01$, significantly higher than has been
commonly assumed for objects in the Kuiper Belt.  Multicolor,
multiepoch photometry shows this pair to have colors typical for the
Kuiper belt with a spectral gradient of 0.35 per 100 nm in the range
between 475 and 775 nm.  Photometric variations at the four epochs we
observed were as large as 12$\pm$3\% but the sampling is insufficient
to confirm the existence of a lightcurve.}

\newpage

\section{INTRODUCTION}

\RZ\ was discovered in images taken on September 8, 1999 at the
Canada-France-Hawaii telescope by Trujillo et al.~(2000).  Its
heliocentric orbit was determined by subsequent observations by this
group and others (Veillet et al.~2000).  The latest orbital elements
reported by the Deep Ecliptic Survey team (Millis et al. 2002), a=44.02
AU, e=0.091, i=0.564 deg, put it in a class with other
low-inclination, low-eccentricity, non-resonant objects sometimes
called ``classical'' Kuiper Belt objects.

The Deep Ecliptic Survey website (Millis et al.~2002) has compiled
astrometric observations of \RZ , mostly obtained in the R band, with a
median R magnitude of 21.7.  More accurate photometric observations
find R=21.5 (Delsanti et al. 2001). The minor planet center, using the
standard asteroidal color conversion, derives an absolute magnitude in
the V band of $H_V$=5.9, suggesting that \RZ\ could be among the larger
of the known transneptunian objects (TNO).  (We show below that \RZ\,
like many TNOs, is significantly redder than the asteroid standard and,
thus, has a fainter $H_V$.)

We identified \RZ\ as a binary in images we took using the Hubble Space
Telescope's (HST) Near-Infrared Camera and Multi-Object Spectrograph
(NICMOS) on 23 April 2003 (Noll and Stephens 2003).  Prompted by this
discovery, a series of followup observations were made with the HST
Advanced Camera for Surveys (ACS) using its High Resolution Camera
(HRC).  We resolved the binary at all four epochs observed by the ACS. 
We have been able to use the relative astrometry available from our HST
images to determine the relative orbit of the binary pair.  With that
comes a determination of the system mass, constraints on albedo and
density, and photometry of the individual components that we describe
in this paper.  Despite its great distance and small size, \RZ\ joins
the ranks of well-characterized solar system objects thanks to its
beneficent duplicity.

\section{OBSERVATION}

\RZ\ has been observed at seven separate epochs with HST as summarized
in Table 1 and Fig.~1.  The binary has been resolved at five of these seven
epochs, and the remaining two provide upper limits.  Three different
instruments were used and we describe these observations separately
below.

\subsection{NICMOS} 

\RZ\ was observed with the NIC2 camera in April 2003 as part of a large
survey of infrared colors of TNOs.  Two 256 second images were obtained
through the F110W filter ($\sim$J) and two 512 second images were
obtained with the F160W filter ($\sim$H).  These were obtained using the
STEP256 MULTIACCUM sequence with the number of samples, NSAMP, being 11
and 12 respectively.  The pixel scale in the NIC2 camera is
approximately 0\arcsec .075 on a side.  At 1.1 and 1.6 $\mu$m the
full-width-half-maximum (FWHM) of the telescope's diffraction-limited
point-spread-function (PSF)is 0\arcsec .097 and 0\arcsec .140.  Thus,
the image is nearly well-sampled at 1.6 $\mu$m and undersampled at 1.1
$\mu$m.  The pair of images in each filter was dithered by 5.5 pixels in
both x and y directions in order to avoid bad pixels, and to potentially
recover some of the spatial information lost in the undersampled image. 

\RZ\ was found in the NICMOS images close to the expected NIC2 default
aperture position.  However, examination of the image immediately
showed that, unlike most of the other objects observed by us, this
object was clearly binary (Noll and Stephens 2003).  The possibility of
confusion between a binary companion and a faint background object was
ruled out by the stability of the relative positions of the pair during
the approximately 100 minutes the object was observed.  Both the proper
motion of the object and parallax contribute to a non-linear apparent
motion of $>$ 3\arcsec .9 compared to fixed objects in the
field.

The four NICMOS images were reduced using standard pipeline calibration
steps.  To obtain relative astrometric positions, each
individual image was fit with a pair of scaled PSFs using an iterative
procedure that minimizes the residuals after subtracting the combined
PSFs.  The derived scale factors give the photometric magnitude of each
component as listed in Table 2.

\subsection{ACS/HRC}

We used the unprecedented power of the ACS/HRC to make followup
observations in August, September and November 2003 (Table 1).  In each
visit we performed a single orbit of observation with two 600 sec
exposures through the F475W filter ($\sim$Sloan g) and two 500 sec
exposures with the F775W filter ($\sim$Sloan i).  The observations were
dithered to minimize the effects of bad pixels and to improve the
sampling in the undersampled F475W images. 

The orbital period was initially unknown so we adopted a two step
strategy to maximize our sampling of the orbit.  Using an assumed
albedo of 5\%, a density of 1000 kg m$^{-3}$, and the observed maximum
separation we derived a minimum period of 10 days.  We noted that other
TNO binaries with known periods, \WW\ and \QT , have periods
substantially longer than the minimum calculated in this way. 
Therefore, we specified that the initial pair of exposures be separated
by 25-50 days.  The remaining two orbits remained on hold until these
data could be analyzed.  After examining the first pair of ACS/HRC
images, we activated the second pair, this time with a specified time
separation of 9-15 days.  As can be seen in Fig.~2, this sampling
stategy not only allowed us to determine the orbital period, but gave
us fortuitously good sampling of the orbit such that we were able to
also constrain the remaining orbital elements as discussed below.  It
should be noted however, that without {\it a priori} knowledge of
the period and with limited control over when observations are
scheduled, sampling is mainly a matter of chance; good sampling will
generally require more observations to adequately constrain an orbit.

Images were reduced using standard HST pipeline calibration.  Dithered
images were combined and cosmic rays removed using the multidrizzle
software package.  We corrected the substantial anamorphic
distortion in the ACS images before deriving relative astrometric
positions.  As with the NICMOS data, we fit the images to a pair of
scalable synthetic PSFs made with version 6.2 of the TinyTim software
package (Krist and Hook 2003).  This procedure also gives us
relative photometry which we normalized to absolute photometry using
the standard HST zeropoints.  An added complication for analysis of the
ACS/HRC data compared to the other instruments is that the sythetic PSFs
are available only in the geometrically-distorted space of the ACS.  We
therefore conducted PSF fitting and subtraction before applying the
distortion correction.

\subsection{STIS}

\RZ\ was observed in November 2001 as part of a program dedicated to
searching for companions to TNOs (c.f.~Brown and Trujillo 2002).  In this
program observations were made in pairs of orbits separated by several
days.  Images were recorded through the clear filter to maximize
sensitivity to faint companions.  Tracking and parallax correction were
turned off and observations were scheduled near times of minimum
apparent motion by the TNO.  The result of this observing strategy is
that the images, including the intermediate frames, are blurred in a
complex pattern dominated by the parallax.  Images can be reconstructed
by modelling the motion and creating blurred PSFs, however this
inevitably results in a loss of sensitivity along the path of apparent
motion.  

The STIS images of \RZ\ consisted of eight 300 sec exposures in each
of two orbits.  Because \RZ\ is relatively bright at R$\sim$21.7, there
is sufficient S/N in each 300 sec integration that they can be examined
individually.  We retrieved these data from the archive using standard
on-the-fly pipeline calibration and examined the images.  The apparent
motion due to changing parallax is variable throughout the orbit and is
at a minimum near the end of the orbits.  Thus, we examined the final
image on each of the two dates without further correction for motion. 
Neither image reveals a
clearly resolved binary, but both can be used to set upper limits on the
separation of approximately 0\arcsec .1.

\section{ANALYSIS}

The images from each epoch have been analyzed separately using a binary
PSF fitting routine that iteratively fits synthetic PSFs on a 0.1 pixel
grid, identifying the best fit at each location through a minimization
of the residuals.  The grid-pair with the lowest minimum is the best
overall fit; we derive the separation, position angle, and relative
flux from this fit.  Synthetic PSFs are generated for each image using
the TinyTim software package (Krist and Hook 2003) with inputs
appropriate for the instrument in use and the location of the image on
the array.  The results from this procedure are found in Tables 1 and 2.

The separations and position angles measured from images at five
epochs, plus the upper limits measured at two epochs, were used as
input into a binary orbit fitting routine.  The program takes fully
into account the time-dependent geometry of the orbit as viewed from
the Earth.  The model searches for a set of orbital elements that
minimize the residuals from the observations and predicted positions. 
Two separate solutions can be found that differ mainly in the
inclination of the orbit plane.  With sufficient observing baseline,
the degeneracy can be broken as the orbit plane angle changes due to
orbital motion of the TNB.  We find one orbit solution has a slightly
lower $\chi ^2$, but cannot yet rule out the mirror solution.  The best
fitting orbit is plotted in Fig.~2.  The resulting orbital elements are
tabulated in Table 3.  

Errors were estimated for each orbital element by two methods; we
report the larger error in each case.  We first used the method of
Lampton et al.~(1976) which involves varying one parameter at a time
and results in a 1-sigma contour.  The second approach holds one orbital
element constant over a range of values while allowing all other
orbital elements to vary freely at each step to minimize $\chi^2$. 
With this method we took the minimum $\chi^2$+1 to be the 1-sigma
contour.  We note that at the times of the two observations in 2001,
the orbit solutions predict separations of s$<$ 0\arcsec .1, in
agreement with the upper limits set by these non-detections.  At all
other dates the difference between the observed and predicted position
is less than 0\arcsec .004.

An important aspect of breaking the orbit degeneracy is determining
when mutual events can occur for this binary system.  For the two
possible orbital solutions, mutual events can occur approximately in
the year 2045 for solution 1 or 2108 for solution 2.  Continued
observations over the next decade can determine which of these
possibilities will occur.

Because of the limited number of observations the orbit solution is
particularly vulnerable to unrecognized sources of error.  Small
changes in the semimajor axis can result in significant changes to the
derived mass.  Despite the fortuitously good sampling of the orbit,
further observations may be able to refine the orbital elements and
should be carried out.

\section {DISCUSSION}

The most direct value of binary orbit determination is that we are able
to derive a mass for the system from direct observation.  For \RZ\ we
find a mass of 3.7 $\pm 0.4 \times 10^{18}$ kg.  This is similar to
the system masses found for two other TNBs, \WW\ and \QT\ which have
been found to be 2.7 $\pm 0.3 \times 10^{18}$ kg (Veillet et al. 2002)
and 3.3 $\pm 3 \times 10^{18}$ kg (Osip et al. 2003).  We note that
the mass of the Pluto-Charon system is some 5000 times greater putting
it into a qualitatively different category than relatively low mass
binaries like \RZ .  

The directly measured mass of \RZ\ is 9-13 times lower than would have
been estimated using the assumptions that have been frequently applied
when estimating mass from photometry, a fact that has global
significance for mass estimates of the Kuiper Belt as a whole.  The MPC
lists an $H_V$ magnitude of 5.9.  If we assume the body(ies) to have a
surface albedo of $p_V$ = 0.04, we derive a diameter of 447 km (for a
single object) or two objects with $d$ = 316 km (for an equal mass
binary).  Making the further assumption that the bulk density is $\rho$
= 1000 kg m$^{-3}$ we would derive masses of 33-47 $\times 10^{18}$ kg.
 Even with a modified assumption of higher albedo, $p_V$ = 0.10, we
would overestimate the mass of this system by factors of 2.3-3.3
depending on whether it was known to be a binary or not.  The
uncertainties in deriving absolute magnitudes in the V band from
ground-based data that are often obtained with different filters for
TNOs that have, as a group, a wide range of non-solar colors, is an important
part of the total error budget.  As shown in Table 4, in the case of
\RZ , the standard color correction produces an overly-bright estimate of
the H$_V$ magnitude.  Global estimates of the mass of the Kuiper Belt
have been made using similar kinds of extrapolations from photometry
and assumptions of low albedo.  In their recent review, Jewitt and Luu
(2000) estimate the mass of objects with diameters 100 km $< d <$ 2300
km is 0.1 M$_{\oplus}$ based on an assumed R-band albedo of $p_R$=0.04.
By contrast, we estimate the R-band albedo of \RZ\ to be $p_R \sim$
0.17, comparable to the very high albedo lower limit found for 28978
Ixion (Altenhoff et al.~2004).  If objects with albedos like \RZ\ are
common, it may be necessary to revise downward mass estimates of the
Kuiper Belt by as much as an order of magnitude.

The mass of the individual binary components cannot be derived from the
relative astrometry we have performed.  It can, in principle, be
derived from absolute astrometric measurements of each component that
would yield the orbit about the systems barycenter.  However, such
measurements are beyond the scope of the observations we have made and
would require a substantial investment of observing time.  If, however,
we make the assumptions that the surface albedos and densities of each
component are identical, a mass ratio can be derived from measured
photometry according to the relation $m_A/m_B = (F_A / F_B)^{3/2}.$
Using the average measured flux in the F475W filter we
obtain $m_A/m_B$ = 1.79.  If the assumption of equal albedos holds, the
individual masses of the binary components are m$_A = 2.39\pm 0.48 \times
10^{18}$ kg and m$_B = 1.34\pm 0.27 \times 10^{18}$ kg.  Similarly, it is
possible to obtain a ratio of the diameters of the two components of
the binary.  Assuming equal albedos, the objects have a diameter ratio
of $d_A/d_B = (F_A / F_B)^{1/2} = 1.21.$  For a density of 1000 kg
m$^{-3}$, the diameters of the A and B components are 166 and 137 km
respectively.  

Once the relative sizes of the objects are assumed known from the
measured relative photometry (and the assumption that each object has
the same albedo) we can plot the albedo, $p$, as a function of density,
$\rho$, from the expression

$$p = F_{\lambda}/F_{\odot} (R \Delta)^2 (4 \pi \rho / 3 m)^{2/3}$$

where R is the sun-object distance, $\Delta$ the earth-object distance,
$F_{\odot}$ the solar flux at 1 AU.  The result for the measured flux,
$F_{\lambda}$,  and derived mass, $m$, of \RZ\ is shown in Fig.~3. 
What is notable is that for densities in the range that would be
expected for the outer solar system, i.e. near a density of 1000 kg
m$^{-3}$, the albedo is greater than 0.1, significantly higher than is
generally assumed for TNOs.  By way of comparison, the albedo of \WW\
for an assumed density of 1000 kg m$^{-3}$ is $p_R$=0.054 (Veillet et
al.~2002) and the average of 9 TNOs with radiometrically determined
diameters is $p_R$ = 0.08 (Altenhoff et al.~2004).

\RZ\ is also notable for the short duration of its orbital period, 12
times shorter than the 574 day period of \WW (Veillet et al. 2002) and 17
times shorter than \QT 's 788 day orbital period (Osip et al.~2003). 
However, this is most likely an artifact of observational bias.  It is
virtually certain that other TNBs will also be found to have comparably
short periods.  Short period objects are of greater interest with
regard to the potential for observable mutual events.  Typical mutual
event seasons for TNBs will be on the order of months (Noll 2003), so
orbits with periods of this order or shorter will maximize mutual event
opportunities.

The eccentricity of TNB orbits is a potentially useful tool to
discriminate between formation models.  In particular, the exchange
reaction model (Funato et al.~2004) predicts that most binaries will
have eccentricities greater than 0.8.  Of the three binary systems that
now have either measured or constrained eccentricity (Veillet et
al.~2002, Osip et al.~2003, this work) only one is
this high.   \RZ\ and \QT\ have mutual orbits
with eccentricities significantly lower than 0.8.  The data available
would suggest that the formation model proposed by Funato et al. is not
the operative mechanism for the majority of binaries formed in the
Kuiper Belt.

The multi-filter photometry we obtained can be used to construct a very
low resolution spectrophotometric spectrum from 0.47 - 1.65 $\mu$m. 
The resultant spectrum is shown in Fig.~4.  The optical color ratios
derived from the F475W and F775W filters result in spectral gradients
of $s$=0.35 and $s$=0.33 for the primary and secondary components
respectively.  We have defined the spectral gradient $s$ as the
fractional change in reflectance per 100 nm, consistent with the
convention introduced by Boehnhardt et al. (2001).  The spectral
gradient is a useful tool for comparing colors obtained with different
filter sets without requiring color transformations.  It implicitly
assumes that a single gradient applies across the optical spectrum for
TNOs, a result that has been confirmed observationally (Jewitt and Luu
2001).  The spectral gradients of 41 non-resonant TNOs measured by us
using WFPC2 (Stephens et al. 2003) range from 0.14 to 0.64.  \RZ\ falls
within the middle of this range.

There is no indication from the data we have compiled that either
primary or secondary component has an intrinsic lightcurve as shown in
Table 4.  On the four occasions we observed with the ACS, the
variations in the normalized absolute flux are within the range
expected from the measurement uncertainties with the exception of one
point on 20 August 2003.  However, this sparse sampling does not rule
out the possibility of intrinsic lightcurve(s) and future observations
should be carried out to investigate this possibility.

\RZ\ joins what will be a growing list of transneptunian binaries with
measured orbits and masses.  The availability of this indispensible
physical information makes these systems among the most interesting
astronomical targets in the solar system.  Their links to the early
history of the solar system mean that they will continue to be the focus
of intensive study as we explore the outer edges of the solar system.

\acknowledgements {Based on observations made with the NASA/ESA Hubble
Space Telescope. These observations are associated with programs \#
9386 and \# 9991.  Support for programs \# 9386 and \# 9991 was
provided by NASA through a grant from the Space Telescope Science
Institute, which is operated by the Association of Universities for
Research in Astronomy, Inc., under NASA contract NAS 5-26555.}

\newpage

{
\null\vskip .1in
\tabskip=1.5em 
\baselineskip=12pt
\tolerance=10000
$$\vbox{ \halign {
\hfil #\hfil & #\hfil & #\hfil & #\hfil & #\hfil & #\hfil \cr
\multispan6\hfil{\bf Table 1: Positional Data}\hfil \cr
\noalign { \vskip 12pt \hrule height 1pt \vskip 1pt \hrule height 1pt \vskip 8pt } 
date &  separation      & pos.~angle& R(AU)& $\Delta$(AU)& instrument\cr
     &(10$^{-3}$\arcsec)& (deg) &      &             &           \cr
\noalign { \vskip 8pt\hrule height 1pt \vskip 8pt }
\noalign {\bigskip } 
2001 Nov 09.37 & $\le$100 & -- & 40.971 & 40.775 & STIS \cr
\noalign {\smallskip}
2001 Nov 12.38 & $\le$100 & -- & 40.971 & 40.827 & STIS \cr
\noalign {\smallskip}
2003 Apr 23.14 & 200$\pm$40 & 116$\pm$4 & 41.060 & 41.551 & NIC2 \cr
\noalign {\smallskip}
2003 Aug 20.61 & 72$\pm$7 & 340.3$\pm$4.3 & 41.081 & 40.072 & ACS/HRC \cr
\noalign {\smallskip}
2003 Sep 15.27 & 227$\pm$2 & 101.6$\pm$1.2 & 41.085 & 40.143 & ACS/HRC \cr
\noalign {\smallskip}
2003 Nov 17.50 & 99$\pm$3 & 40.7$\pm$3.2 & 41.096 & 40.983 & ACS/HRC \cr
\noalign {\smallskip}
2003 Nov 29.30 & 74$\pm$2 & 194.6$\pm$5.6 & 41.098 & 41.189 & ACS/HRC \cr
\noalign {\smallskip}
\noalign {\vskip 8pt \hrule height 1pt }  
  } }$$}

{
\null\vskip .1in
\tabskip=1.5em 
\baselineskip=12pt
\tolerance=10000
$$\vbox{ \halign {
\hfil #\hfil & #\hfil & #\hfil & #\hfil & #\hfil & #\hfil  \cr
\multispan6\hfil{\bf Table 2: Photometry}\hfil \cr
\noalign { \vskip 12pt \hrule height 1pt \vskip 1pt \hrule height 1pt \vskip 8pt } 
date &  filter & pivot               & component A & component B & phase \cr
     &           & $ \lambda (\mu$m) & (mag)       & (mag)       & (deg) \cr
\noalign { \vskip 8pt\hrule height 1pt \vskip 8pt }
\noalign {\bigskip } 
2003 Apr 23.14 & F110W & 1.1285 & 20.69$\pm$0.02 & 21.18$\pm$0.02 & 1.2 \cr
               & F160W & 1.6060 & 20.21$\pm$0.02 & 20.65$\pm$0.02 & \cr
\noalign {\medskip}
2003 Aug 20.61 & F475W & 0.47757 & 23.46$\pm$0.03 & 23.78$\pm$0.05 & 0.1 \cr
               & F775W & 0.76651 & 21.64$\pm$0.02 & 21.89$\pm$0.03 & \cr
\noalign {\smallskip}
2003 Sep 15.27 & F475W  &         & 23.39$\pm$0.02 & 23.80$\pm$0.04 & 0.5 \cr
               & F775W  &         & 21.64$\pm$0.02 & 22.01$\pm$0.02 & \cr
\noalign {\smallskip}
2003 Nov 17.50 & F475W  &         & 23.54$\pm$0.05 & 24.07$\pm$0.06 & 1.4 \cr
               & F775W  &         & 21.83$\pm$0.03 & 22.22$\pm$0.04 & \cr
\noalign {\smallskip}
2003 Nov 29.30 & F475W  &         & 23.47$\pm$0.05 & 23.89$\pm$0.05 & 1.4 \cr
               & F775W  &         & 21.88$\pm$0.03 & 22.20$\pm$0.04 & \cr
\noalign {\smallskip}
\noalign {\vskip 8pt \hrule height 1pt }  
  } }$$}

{
\null\vskip .1in
\tabskip=1.5em 
\baselineskip=12pt
\tolerance=10000
$$\vbox{ \halign {
#\hfil & #\hfil & #\hfil   \cr
\multispan3\hfil{\bf Table 3: Derived Orbital Parameters}\hfil \cr
\noalign { \vskip 12pt \hrule height 1pt \vskip 1pt \hrule height 1pt \vskip 8pt } 
element & orbit solution 1   & orbit solution 2 \cr
        &    & \cr
\noalign { \vskip 8pt\hrule height 1pt \vskip 8pt }
\noalign {\bigskip } 
period (days)        & 46.263$\pm {0.006\atop 0.074}$& 46.233$\pm {0.006\atop 0.065}$\cr
a (km)               & 4,660$\pm$170                 & 4,700$\pm$170    \cr
e                    & 0.460$\pm$ 0.013              & 0.454$\pm$ 0.013 \cr
i (deg)              & 152$\pm$3                     & 51$\pm$ 3        \cr
periapse (JD-2450000)& 1859$\pm$2                    & 1860$\pm$ 2      \cr
$\theta$ (rad)       & 0.80$\pm$0.13                 & 1.27$\pm$ 0.09   \cr
w (rad)              & 2.79$\pm$0.07                 & 2.92$\pm$ 0.06   \cr
\noalign {\smallskip}
$\chi ^2$            & 1.18                          & 1.21             \cr
\noalign {\smallskip}
mutual events (year) & $\sim$2045                    & $\sim$2108       \cr
\noalign {\smallskip}
system mass (10$^{18}$ kg) & 3.7$\pm$0.4             & 3.8$\pm$0.4      \cr
\noalign {\bigskip } 
\noalign {\vskip 8pt \hrule height 1pt }  
  } }$$}

{
\null\vskip .1in
\tabskip=1.5em 
\baselineskip=12pt
\tolerance=10000
$$\vbox{ \halign {
\hfil #\hfil & \hfil #\hfil & #\hfil & \hfil #\hfil & #\hfil  \cr
\multispan5\hfil{\bf Table 4: Absolute magnitudes}\hfil \cr
\noalign { \vskip 12pt \hrule height 1pt \vskip 1pt \hrule height 1pt \vskip 8pt } 
date &  component & filter & H$_{\lambda}$ (mag)  & normalized \cr
\noalign { \vskip 8pt\hrule height 1pt \vskip 8pt }
\noalign {\bigskip } 
2003 Aug 20.61 & A & F475W &  7.36$\pm$0.03 & 0.88$\pm$0.03 \cr
2003 Sep 15.27 &   &       &  7.23$\pm$0.03 & 0.99$\pm$0.02 \cr
2003 Nov 17.50 &   &       &  7.20$\pm$0.07 & 1.02$\pm$0.06 \cr
2003 Nov 29.30 &   &       &  7.12$\pm$0.07 & 1.10$\pm$0.07 \cr
average        &   &       &  7.22$\pm$0.03 &               \cr
\noalign {\smallskip}
2003 Aug 20.61 & B &       &  7.68$\pm$0.05 & 0.96$\pm$0.04 \cr
2003 Sep 15.27 &   &       &  7.64$\pm$0.05 & 1.01$\pm$0.04 \cr
2003 Nov 17.50 &   &       &  7.73$\pm$0.08 & 0.93$\pm$0.07 \cr
2003 Nov 29.30 &   &       &  7.54$\pm$0.07 & 1.10$\pm$0.07 \cr
average        &   &       &  7.64$\pm$0.03 &               \cr
\noalign {\smallskip}
2003 Aug 20.61 & A & F775W &  5.54$\pm$0.02 & 0.97$\pm$0.02 \cr
2003 Sep 15.27 &   &       &  5.48$\pm$0.03 & 1.03$\pm$0.03 \cr
2003 Nov 17.50 &   &       &  5.49$\pm$0.05 & 1.02$\pm$0.05 \cr
2003 Nov 29.30 &   &       &  5.53$\pm$0.05 & 0.98$\pm$0.05 \cr
average        &   &       &  5.51$\pm$0.02 &               \cr
\noalign {\smallskip}
2003 Aug 20.61 & B &       &  5.79$\pm$0.03 & 1.05$\pm$0.03 \cr
2003 Sep 15.27 &   &       &  5.85$\pm$0.03 & 0.99$\pm$0.03 \cr
2003 Nov 17.50 &   &       &  5.88$\pm$0.06 & 0.97$\pm$0.05 \cr
2003 Nov 29.30 &   &       &  5.85$\pm$0.06 & 0.99$\pm$0.05 \cr
average        &   &       &  5.84$\pm$0.02 &               \cr
\noalign {\smallskip}
\noalign {\vskip 8pt \hrule height 1pt }  
  } }$$}

\begin{figure}
\includegraphics[totalheight=0.4\textheight,angle=0]{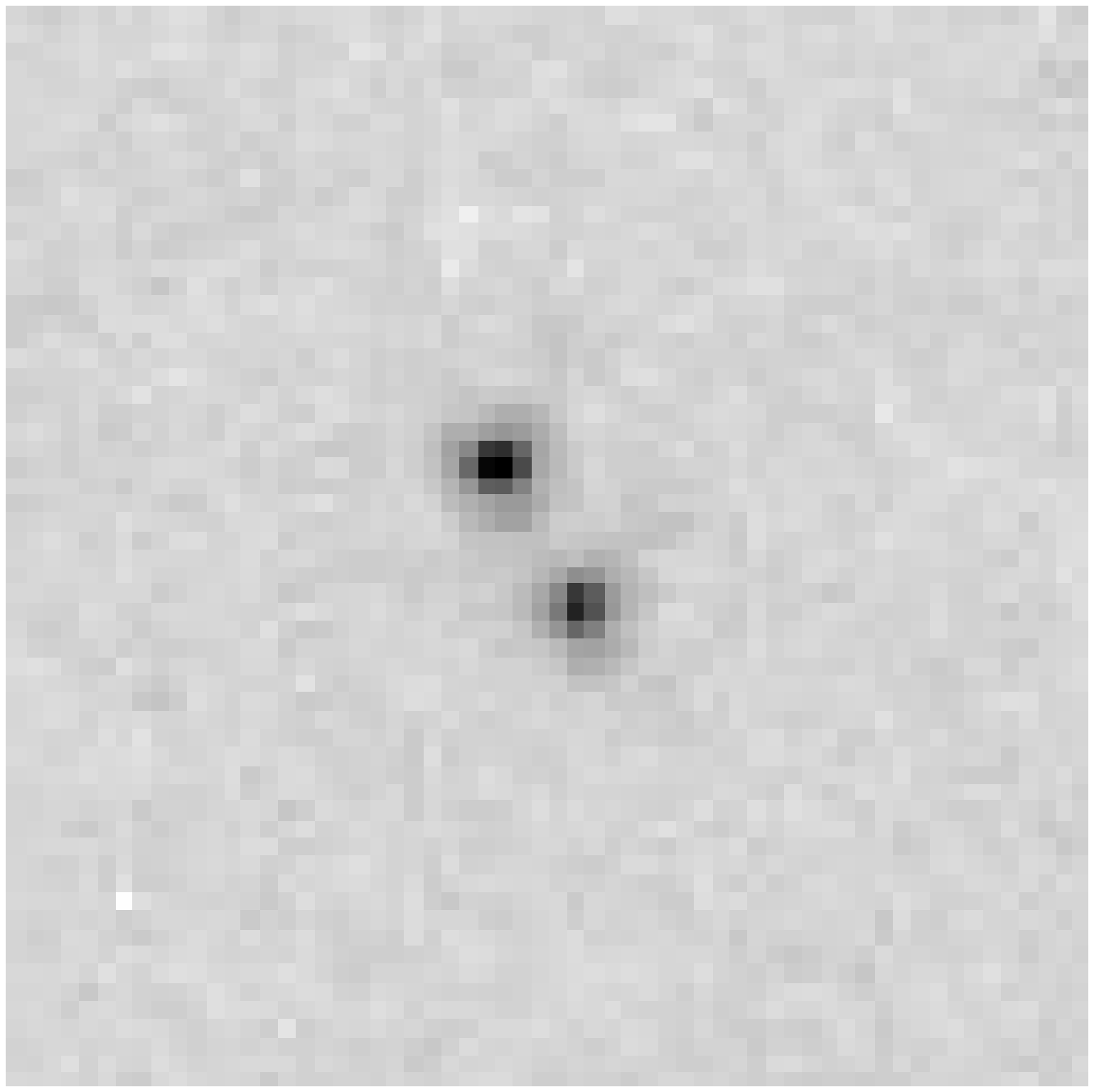}
\includegraphics[totalheight=0.4\textheight,angle=0]{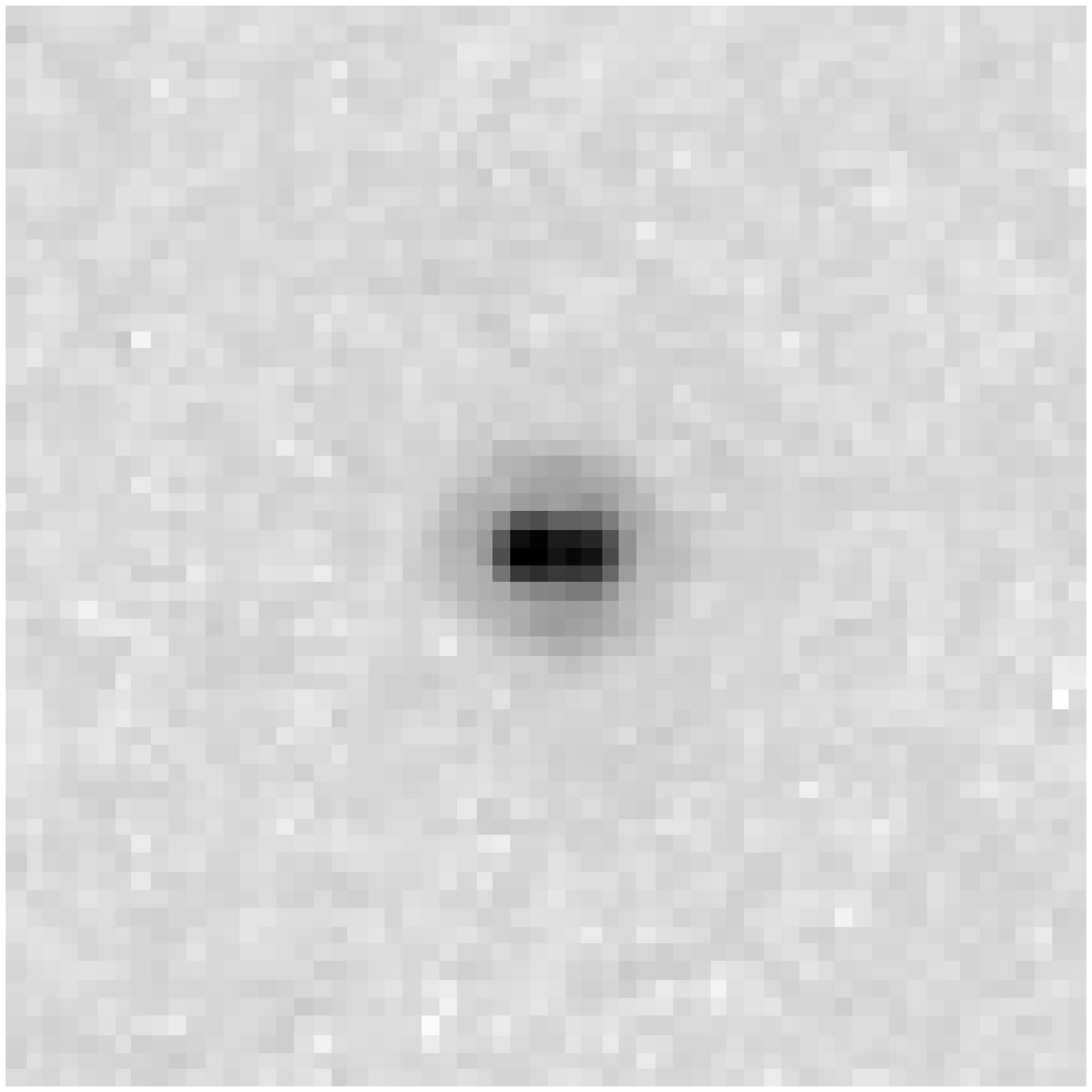}
\includegraphics[totalheight=0.4\textheight,angle=0]{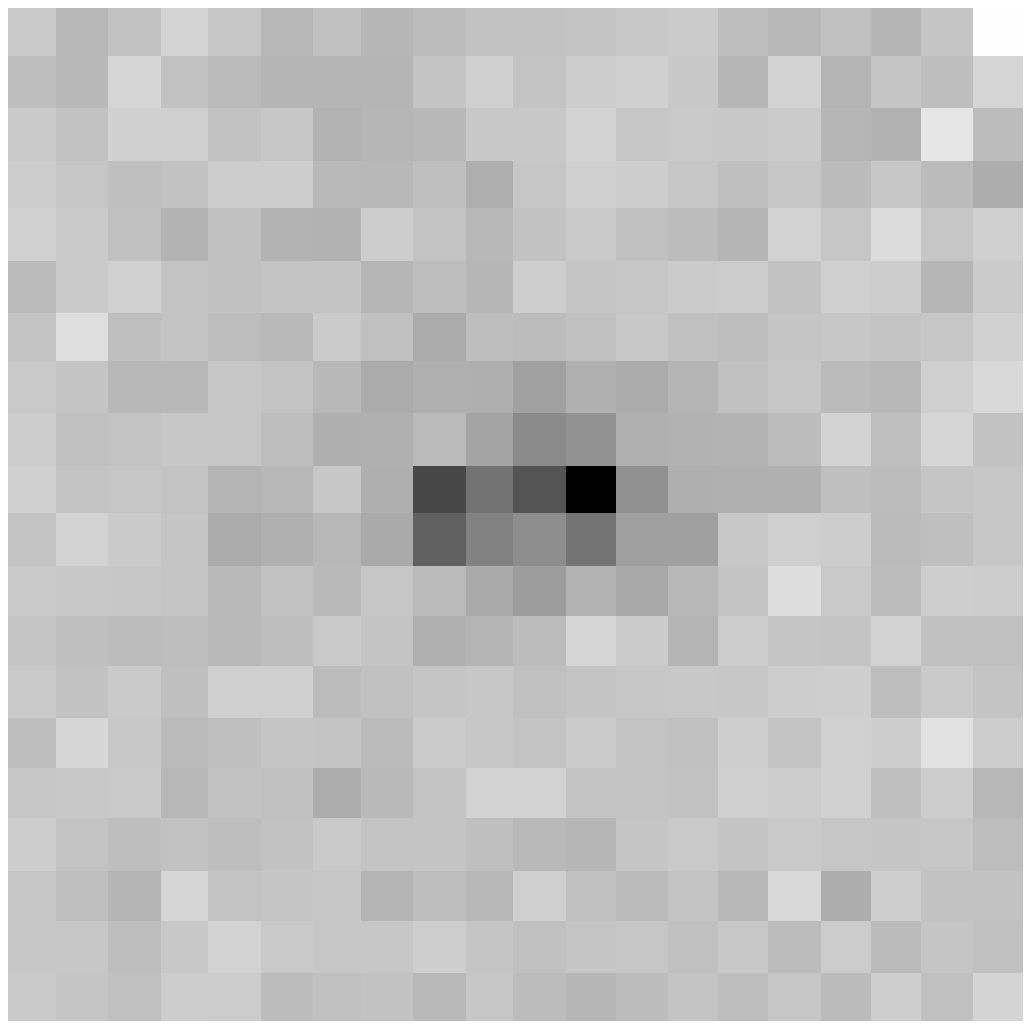}
\includegraphics[totalheight=0.4\textheight,angle=0]{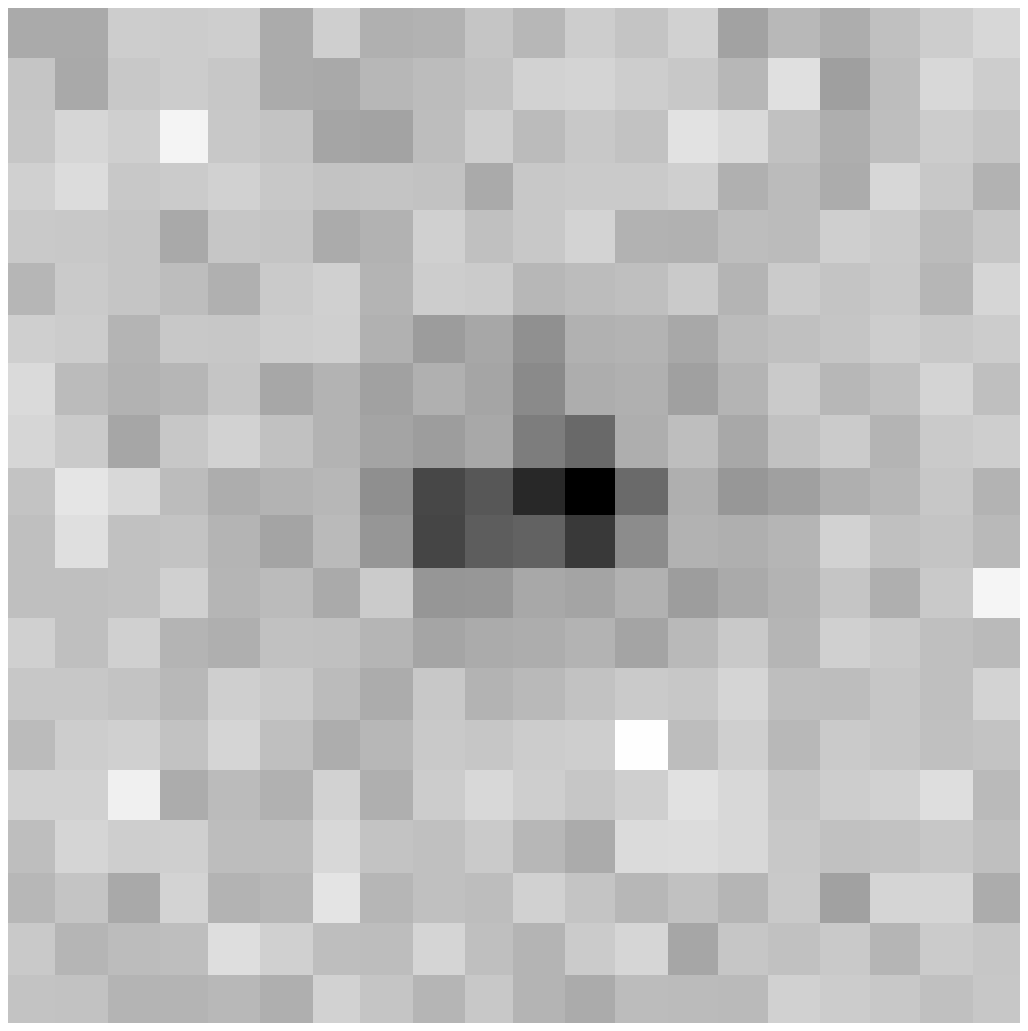}

\caption {Images of \RZ\ obtained with a) ACS using the F475W filter on
15 September 2003, b) ACS using the F775W filter on 20 August 2003, c) NICMOS
using the F110W filter on 23 April 2003, and d) NICMOS using the F160W
filter on 23 April 2003.  Each image is 1.5 arcsec on a side.  The
orientations are in native detector coordinates; the position angles
listed in Table 1 are calculated using the spacecraft orientation information
available in the data files.
\label{fig1a}} 
\end{figure}

\begin{figure}
\includegraphics[totalheight=0.5\textheight,angle=0]{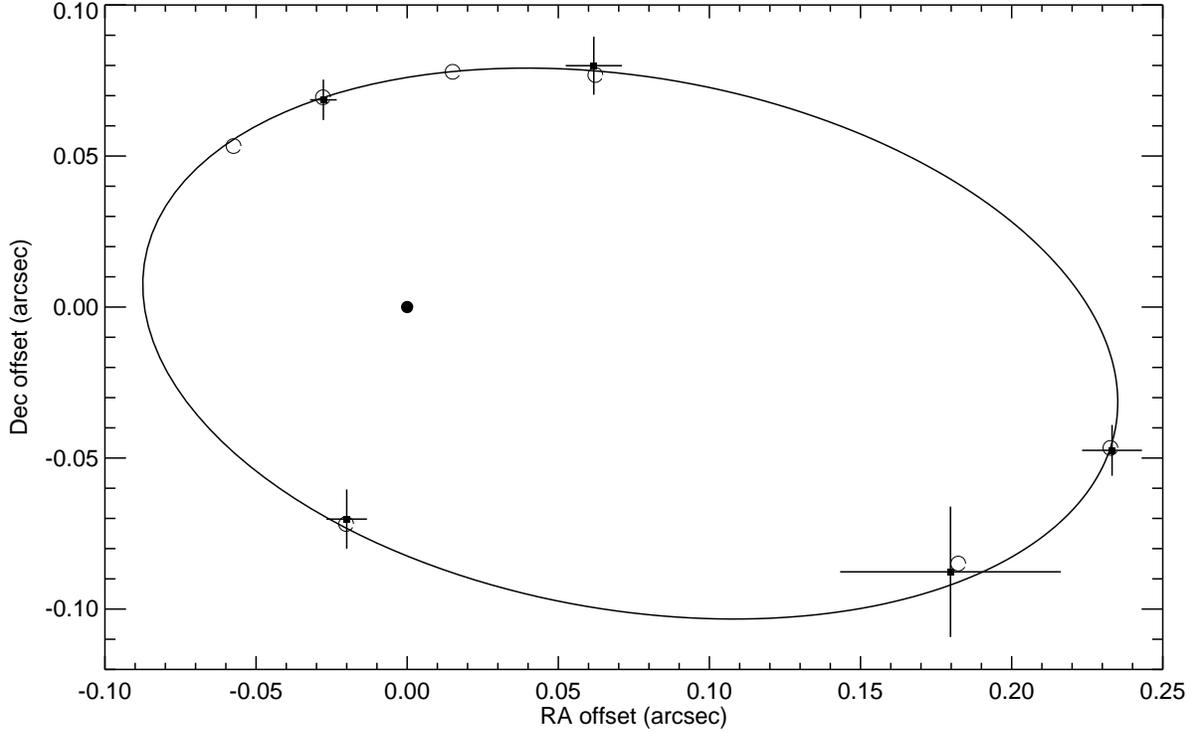}

\caption {The relative positions of the fainter of the two components
of \RZ\ relative to the brighter component are shown with positional
uncertainties indicated.  The curve is the best fit orbit as determined
by our orbit fitting procedure at an arbitrarily selected instant in
the interval covered by our data.  Because of the changing viewing
geometry and the precession of the orbit plane, the apparent shape and
size of the orbit change slightly with time.  The predicted position of
the secondary at each of the seven epochs is indicated with a small
circle; the small deviations of these circles from the orbit ellipse
are due to the aforementioned changes in viewing geometry.  The two
circles without corresponding measurements indicate the predicted
position of the secondary at the time of the STIS-derived upper limits.
\label{fig2}}  \end{figure}

\begin{figure}
\includegraphics[totalheight=0.6\textheight,angle=0]{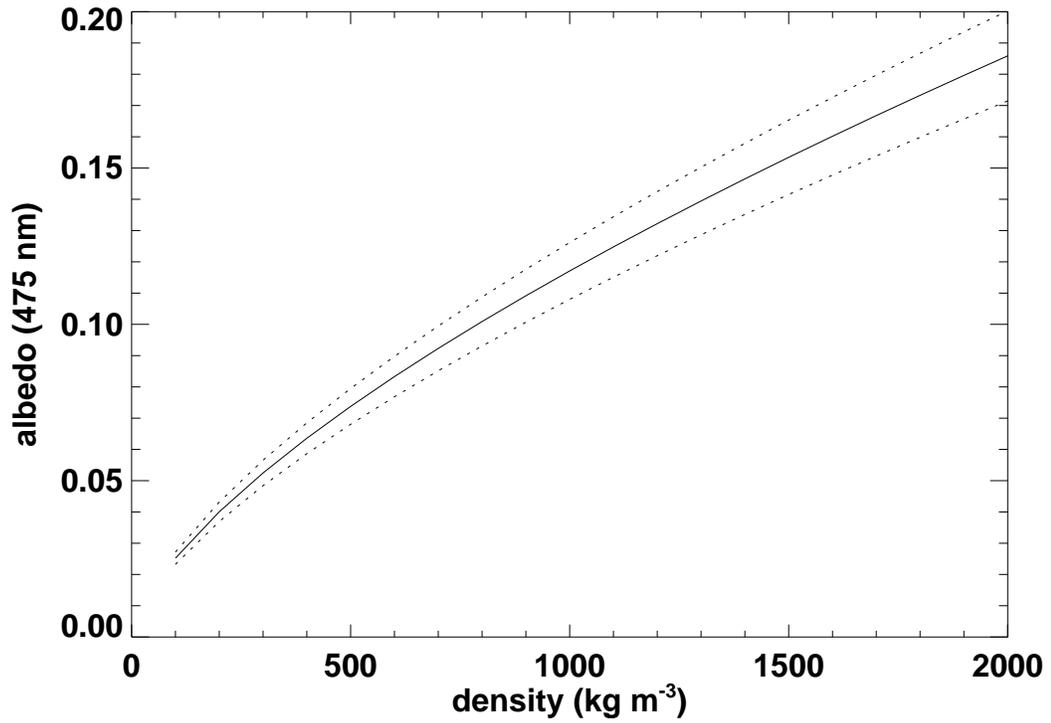}

\caption {Plot of albedo at 475 nm as a function of assumed density. 
The solid curve is computed using the average flux measured at 475 nm
with the uncertainties bounded by the dotted curves. 
For a density of $\rho$ = 1000 kg m$^{-3}$, the average albedo is
$p_{475}$ = 0.12, higher than generally assumed for transneptunian
objects.  If albedos as high as this are typical, the mass of the
Kuiper Belt may be currently overestimated by an order of magnitude. 
\label{fig3}} \end{figure}

\begin{figure}
\includegraphics[totalheight=0.6\textheight,angle=0]{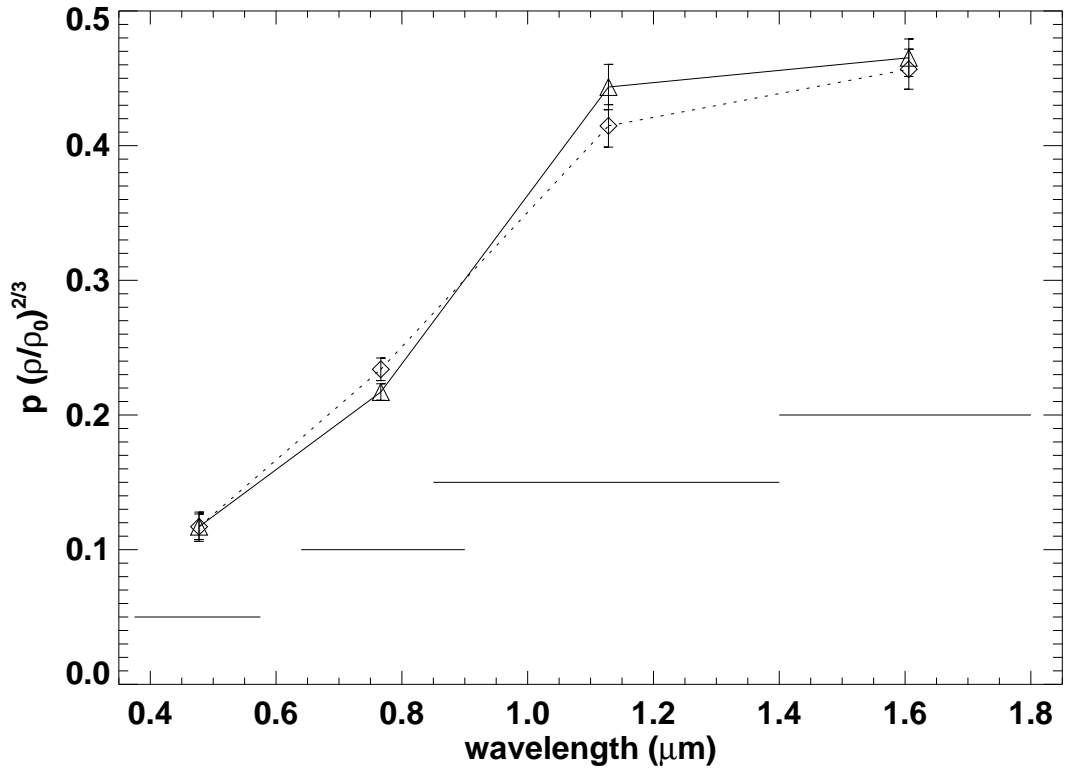}

\caption {The albedo of each component of \RZ\ is plotted at each of
four wavelengths measured.  The curves shown assume that both
components have the same albedo at 475 nm and share a bulk density of
$\rho_0$ = 1000 kg m$^{-3}$. The albedo scales with density to the 2/3 power.  The
solid curve corresponds to the larger member of the binary, component A,
and the dotted curve to the smaller component B.  The solid bars at the
bottom of the figure are show the bandpasses of the four filters used.
\label{fig4}} \end{figure}

\end{document}